\documentclass[twocolumn,showpacs,preprintnumbers, superscriptaddress, amsmath,amssymb, prl, reprint]{revtex4-1}
\usepackage{amsmath}
\usepackage{stmaryrd}
\usepackage{txfonts}
\usepackage{amssymb}
\usepackage{mathrsfs}
\usepackage{graphicx}
\usepackage{dcolumn}
\usepackage{bm}
\usepackage{epsfig}
\usepackage{braket}
\usepackage{color}
\usepackage{hyperref}
\usepackage{perpage} 
\MakePerPage{footnote} 
\begin{document}
\preprint{\href{http://dx.doi.org/10.1103/PhysRevLett.111.166602}{S. -Z. Lin and C. D. Batista, Phys. Rev. Lett. {\bf 111}, 166602 (2013).}}

\title{Orbital Magnetism Induced by Heat Currents in Mott insulators }

\author{Shi-Zeng Lin}
\affiliation{Theoretical Division, T4, Los Alamos National Laboratory, Los Alamos, New Mexico 87545, USA}

\author{Cristian D. Batista}
\affiliation{Theoretical Division, T4, Los Alamos National Laboratory, Los Alamos, New Mexico 87545, USA}

\begin{abstract}
We derive the effective heat current density operator for the strong-coupling  regime of Mott insulators. Similarly to the case of the electric current density, 
the leading contribution to this effective operator is proportional to the local scalar spin chirality $\hat{\chi}_{jkl}=  \mathbf{S}_l\cdot\left(\mathbf{S}_j\times \mathbf{S}_k\right)$. 
This common form of the effective heat and electric current density operators leads to a novel cross response in Mott insulators. A heat current 
induces a distribution of orbital magnetic moments in systems  containing loops of  an odd number of hopping terms. The relative orientation of the orbital moments
depends on the particular lattice of magnetic ions. This subtle effect arises
from the symmetries that the heat and electric currents have in common. 
\end{abstract}
\date{\today}
\maketitle

Mott insulators play a pivotal role in condensed matter physics. Besides being ``parent sates'' of exotic emergent phenomena, such as high-$T_c$ superconductivity, 
they are  the source of most  of the known insulating quantum magnets. The minimal and paradigmatic model for describing the electronic degrees of freedom of  Mott insulators is the half-filled (one electron per atom) single-band Hubbard Hamiltonian. This model includes an intra-atomic Coulomb interaction $U$ and a kinetic energy term that allows to move electrons between different atoms with a hopping amplitude $t$. Electrons become strongly localized in the limit $U \gg t$ because of the very high Coulomb energy barrier for double occupying an atomic orbital. Consequently, the low-energy physics of strongly coupled Mott insulators can be entirely described in terms of the remaining  spin degree of freedom.

The formal procedure for reducing the original Hubbard model to an effective spin Hamiltonian is a canonical transformation plus a projection into the lowest 
energy subspace which is  adiabatically connected with the subspace of states containing exactly one electron per atom in the $t \to 0$ limit.
In this way, one can derive the  well-known  Heisenberg Hamiltonian that was originally introduced as a phenomenological model for quantum ferromagnets.~\cite{Heisenberg28}
However, a low-energy physics which is entirely describable in terms of spin degrees of freedom does not imply that  charge degrees of freedom are completely frozen. In fact, the effective antiferromagnetic (AFM) exchange between local moments arises from the combination of a finite electronic  localization length and the fermionic statistics.

As it was pointed out in  Ref.~\onlinecite{Bulaevskii2008}, the finite localization length can also lead to non-uniform charge  distributions or electric orbital currents. Indeed, both phenomena can occur in equilibrium if the Mott insulator undergoes a symmetry breaking phase transition.
Charge redistributions arise from states that spontaneously break the equivalence between bonds~\cite{Kamiya12}, while orbital currents emerge in states that exhibit spontaneous  scalar spin chirality, $\langle\hat{\chi}_{jkl}\rangle=\langle\mathbf{S}_l\cdot\left(\mathbf{S}_j\times \mathbf{S}_k\right)\rangle\neq 0$,~\cite{Khaled09,Chubukov13} The notion of scalar spin chirality appears in numerous discussions of magnets and superconductors,~\cite{Wen1989,Shastry1990,Sulewski1991,Kawamura1992,Momoi1997,Kawamura1998,Taguchi2001,Rosler2006,Schweika2007} and its identification
with an observable (electric current density) is crucial for measuring this subtle order parameter.

The electric charge and current density operators, $\rho$ and ${\bf I}^{(c)}$, are related by the continuity equation that reflects the conservation of the total charge. Similarly, energy conservation leads to a second continuity equation for the energy 
and heat current density operators $\epsilon$ and ${\bf I}^{(h)}$. The only symmetry operation that distinguishes $\rho$ and ${\bf I}^{(c)}$ from $\epsilon$ and ${\bf I}^{(h)}$ is 
charge conjugation. This simple observation leads to subtle connection between the  effective electric and heat current density operators and  the main physical consequence is the central result of this Letter.

We derive the effective heat current density operator which is also proportional to the local scalar spin chirality. This common nature
of the effective electric and heat current density operators leads to a novel effect in Mott insulators:  DC heat current produced by a temperature gradient can induce an array of orbital magnetic moments, which is different from the spin  ordering. Although the problem of heat conduction in Mott insulators was investigated for many years, \cite{Yoffa1975,Yoffa1977} we are not aware of any study of  heat current-induced orbital magnetic moments. Our predictions can be tested by  performing nuclear magnetic resonance (NMR)  measurements in presence of a finite temperature gradient.

We start by considering a half-filled single-band Hubbard model defined on an arbitrary lattice
\begin{equation}\label{eq2}
\mathcal{H}=-\sum _{ j k,\sigma } t_{jk} \left(c_{j\sigma }^{\dagger }c_{k\sigma } + c_{k\sigma }^{\dagger }c_{j\sigma }\right)+\frac{U}{2}\sum _j\left(n_{j}-1\right)^2,
\end{equation}
 where the operator $c_{j\sigma }^{\dagger }$ ($c_{j\sigma }$) creates (annihilates) an electron with spin $\sigma=\uparrow, \downarrow$ on the atom $j$  and $t_ {jk}$ denotes the hopping between the $j$ and $k$ atoms. $n_j=\sum_{\sigma}n_{j\sigma}\equiv\sum_{\sigma}c_{j\sigma }^{\dagger }c_{j\sigma }$ is the electron number operator for the atom $j$. 

The Hamiltonian $\mathcal{H}$ has $2^N$ degenerate ground states  for $t_{jk}=0$ ($N$ is the total number of atoms in the lattice) because the spin of the electron that occupies each atom or site  can either be up or down. States in this ground space will be denoted by $\Ket{\phi}$. The massive degeneracy is lifted for finite $t_{jk}/U\ll 1$ and the new low-energy 
eigenstates, $\Ket{\psi}$, can be obtained by applying a unitary transformation to the states $\Ket{\phi}$:  $\Ket{\psi}=\exp(-\mathcal{S})\Ket{\phi}$. ${\cal S}$ is the (antihermitian) generator of the unitary transformation. The effective low-energy operator ${\tilde {\cal O}}$ for a given observable ${\cal O}$ is obtained by projecting it  into the low-energy subspace spanned by the states $\Ket{\psi}$. However, in order to  express ${\tilde {\cal O}}$ as a function of spin operators only, it is necessary to work in the basis of $\Ket{\phi}$ states. In this basis we have
$
\tilde {\cal O} = e^\mathcal{S} P_{\psi}  {\cal O} P_{\psi} e^\mathcal{-S} = P_{\phi} e^\mathcal{S}  {\cal O} e^{-\mathcal{S}} P_{\phi},
$
where $ P_{\psi} = \exp(-\mathcal{S}) P_{\phi} \exp(\mathcal{S}) $ is the projector on the subspace generated by the states $\Ket{\psi}$, while $P_{\phi}$ projects on the subspace generated  by the the singly-occupied states $\Ket{\phi}$. 
For  ${\cal O} = \mathcal{H}$ we obtain the effective AFM Heisenberg spin Hamiltonian,
\begin{equation}\label{eq3}
\tilde{\mathcal{H}}=\sum _{\langle ij\rangle}J_{jk}\left(\mathbf{S}_j\cdot\mathbf{S}_k-\frac{1}{4}\right),
\end{equation}
with $J_{jk}=4 t_{jk}^2/U$, $\mathbf{S}_j =\sum_{\mu,\nu} c_{j\mu}^{\dagger} {\boldsymbol \sigma}_{\mu\nu} c_{j\nu}
$ and ${\boldsymbol \sigma}$ is the vector of Pauli matrices. 
In a similar way, we can obtain the effective operators for the charge $\rho_j=e\sum_{\sigma}c_{j\sigma }^{\dagger }c_{j\sigma } $ and electric current density
\begin{equation}\label{eq5}
{\bf I}_{jk}^{(c)}=-\frac{i e }{\hbar }\sum _{\sigma } t_{jk} \left(c_{k\sigma }^{\dagger }c_{j\sigma}-c_{j\sigma }^{\dagger }c_{k\sigma}\right)\hat{e}_{jk},
\end{equation}
where $\hat{e}_{jk}$ is a unit vector along the bond $jk$.~\cite{Bulaevskii2008}  Here we are using the linear dimension of the unit cell as our unit of length. These two operators are related by the continuity equation on the lattice,
$\partial_t \rho +\nabla\cdot \mathbf{I}^{(c)}$=0, 
that arises from the conservation of the total number of electrons $[\mathcal{H},\ \sum_j n_j ]=0$, with $n_j =\sum_{\sigma} c_{j\sigma }^{\dagger }c_{j\sigma} $. Because the smallest loop in a lattice  is a triangle, contributions to the effective current density operator must involve at least three spins. In addition,  the electric current density is a scalar under spin rotations and odd under timer reversal. Therefore, three spin ($jkl$) contributions must be proportional to the scalar spin chirality $\hat{\chi}_{jkl}$~\cite{Bulaevskii2008}:
\begin{equation}
\label{effcc}
\tilde{{\bf I}}_{jk}^{(c)}=\frac{e}{\hbar} \hat{e}_{jk} \sum_{l} \gamma_{jkl}  \mathbf{S}_l\cdot\left(\mathbf{S}_j\times \mathbf{S}_k\right),
\end{equation}
where $\gamma_{jkl}=-24{t_{jk}t_{kl}t_{lj}}/{U^2}+\mathcal{O}(t^5/U^4)$. $\rho_j$ is a scalar under spin rotations and even under time reversal. 
Therefore, three-spin contributions ($jkl$) must consist of  a linear combination of  scalar products of two spin operators~\cite{Bulaevskii2008}:
\begin{equation}
\label{eq6}
\tilde{\rho}_j=e + e \sum_{kl}  \beta_{jkl}  \left(\mathbf{S}_j\cdot \mathbf{S}_k+\mathbf{S}_j\cdot\mathbf{S}_l-2 \mathbf{S}_k\cdot \mathbf{S}_l\right),
\end{equation}
with $\beta_{jkl}={8 t_{jk}t_{kl}t_{lk}}/{U^3}+\mathcal{O}(t^4/u^4)$. The sum of the prefactors in front
of each of the three scalar products must be equal to zero because of the Pauli exclusion principle: $\tilde{\rho}_j=e$ on a triangle of three fully polarized spins.

\begin{figure}[t]
\includegraphics[width=\columnwidth]{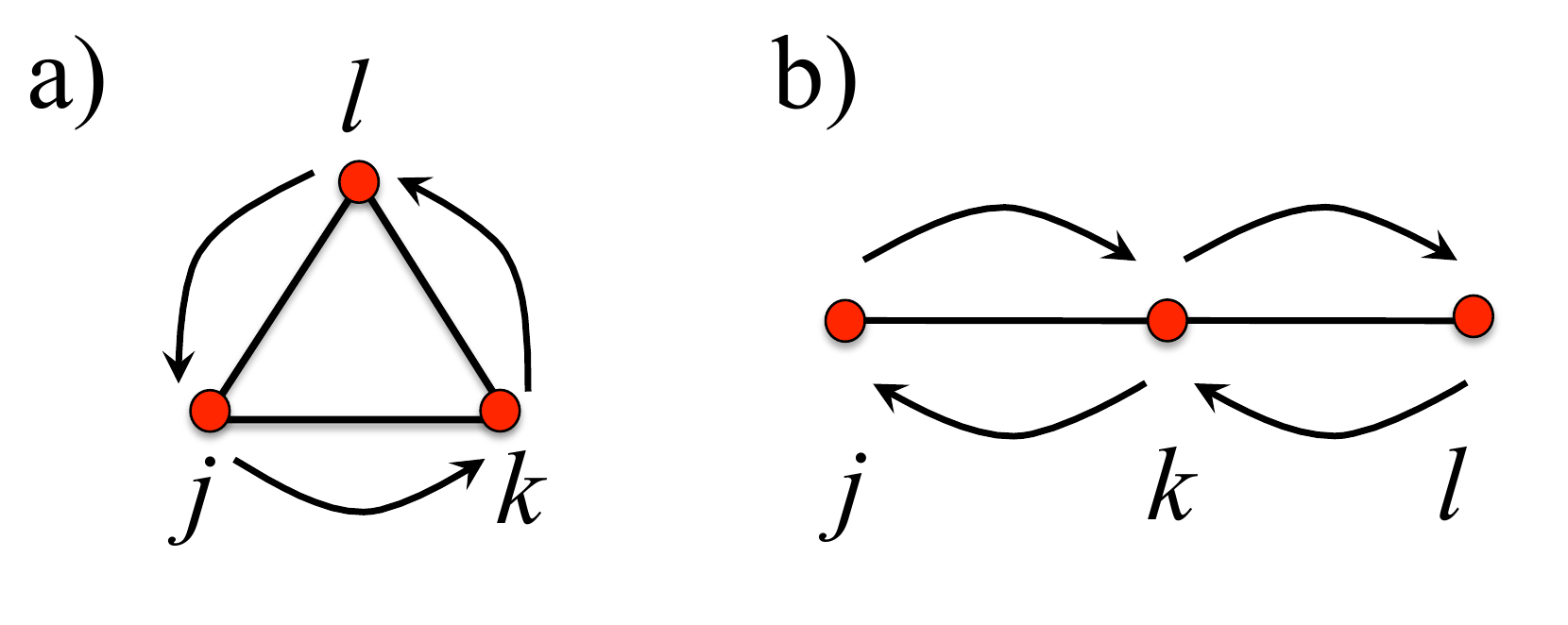}
 \caption{(color online) Leading order contributions to the effective (a) electric and (b) heat current density operators. }
\label{f1}
\end{figure}

It is interesting to note that both $\tilde{\rho}_j$ and ${\tilde {\bf I}}_{jk}^{(c)}$ are odd in the hopping amplitudes. The reason is that charge conjugation (particle-hole transformation) changes the sign of the hopping amplitudes ($t_{jk} \to -t_{jk}$) in Eq.~\eqref{eq2}. In other words, because $\rho_j$ and ${\bf I}_{jk}^{(c)}$ are odd under charge conjugation,
contributions to the corresponding effective operators must be odd in the hopping amplitudes. This observation implies that contributions to these effective operators can only come  from loops of an odd number of hopping terms (see Fig.~\ref{f1}). In the effective Heisenberg model description \eqref{eq3},  these are loops of an odd number AFM exchange interactions. Therefore, geometric frustration is a precondition for having non-trivial effective charge and electric current density operators in Mott insulators.

 Equation~\eqref{eq6} implies that magnetic configurations which break the equivalence between different bonds lead to electric charge redistributions. This simple observation has multiple consequences. For instance, the  charge redistribution induced by certain spin orderings can lead to a net electric polarization.~\cite{Bulaevskii2008} This magnetically driven ferroelectricity is observed in type-II multiferroic materials and Eq.~\eqref{eq6}  allows to compute the electronic contribution to the electric polarization.\cite{Cheong07} For example, the charge effects that have been recently observed in the Mott insulator Cu$_3$MoO$_9$ can be explained by applying this equation.~\cite{Kuroe11,Matsumoto12}   Topological defects provide another example of spin configurations that typically break the equivalence between bonds. According to Eq.~\eqref{eq6}, if the underlying spin model is frustrated, this defects must induce an electric charge redistribution. This observation was recently exploited by D. Khomskii to 
 demonstrate that magnetic monoples in spin ice carry a net electric dipole. \cite{Khomskii12}

After introducing the effective charge and current density operators, we are ready to connect the latter one with the effective heat current density operator.
The total energy  is conserved by $\mathcal{H}$ because $\partial_t \mathcal{H}=0$, and this conservation law is expressed by a second continuity equation:
$\partial_t \epsilon_j+\nabla\cdot \mathbf{I}^{(h)}$=0. $\epsilon_j$ is the energy density and 
\begin{equation}
{\bf I}_{jk}^{(h)} =  - \frac{t_{jk}}{2}{{\hat{e}}_{jk}}\sum_\sigma  (c_{j\sigma }^\dagger {\dot c_{k\sigma }} - \dot c_{j\sigma }^\dagger {c_{k\sigma }} - c_{k\sigma }^\dagger {\dot c_{j\sigma }} + \dot c_{k\sigma }^\dagger {c_{j\sigma }}),
\label{eq8}
\end{equation}
is the heat current  density~\cite{Paul2003}. The time derivative of the creation and annihilation operators is obtained from the Heisenberg equation
$
{\dot c}_{j\sigma}^{(\dagger)} = i [\mathcal{H}, {c}_{j\sigma}^{(\dagger)} ]/\hbar.
$
By replacing ${\dot c}_{j\sigma}^{(\dagger)}$ in  Eq.~\eqref{eq8}, we obtain the following contributions to the heat current density operator ${\bm I}^{(h)}_{jk} = {\bm I}^{(h,U)}_{jk}  + {\bm I}^{(h,t)}_{jk} $
\begin{eqnarray}
{\bm I}^{(h,U)}_{jk} &=& \frac{it_{jk}\hat{e}_{jk}}{2\hbar} U  \delta\rho_{jk}  \sum_{\sigma} [c^{\dagger}_{k \sigma}     c_{j \sigma} -c^{\dagger}_{j \sigma}  c_{k\sigma}  ],
\nonumber \\
{\bm I}^{(h,t)}_{jk} &=&\frac{it_{jk}  \hat{e}_{jk} }{2\hbar}  \sum_{ l \sigma}  (t_{lk} c^{\dagger}_{l \sigma}   c_{j \sigma}+ t_{jl} c^{\dagger}_{k \sigma}  c_{l \sigma} - {\rm H. c.} ),
\label{heatc2}
\end{eqnarray}
where $\delta\rho_{jk}  = n_j  + n_k  - 2$. 

The electric and heat current operators have the same symmetry properties except for the parity under charge conjugation (both $\epsilon_j$ and ${\bm I}^{(h)}_{jk} $ 
are even under charge conjugation). Therefore, the leading order contribution to the effective heat current density operator,  from trimers  containing the bond $jk$, is also proportional to the scalar spin chirality $\hat{\chi}_{jkl}$. However, the proportionality constant must be {\it even} in the hopping amplitude.
By performing the canonical transformation for ${\cal O}={\bm I}^{(h)}_{jk}$, we obtain
\begin{equation}
\label{effhc}
\tilde{{\bf I}}_{jk}^{(h)}=\frac{1}{\hbar} \hat{e}_{jk}  \sum_{l,l'} \left(\alpha_{j,kl}    \mathbf{S}_l+\alpha_{k,jl'}    \mathbf{S}_{l'}\right)\cdot\left(\mathbf{S}_j\times \mathbf{S}_k\right),
\end{equation}
with $\alpha_{j,kl}=-{8 t_{jk}^2 t_{kl}^2}/{U^2}$ and $\alpha_{k,jl'}=-{8 t_{jk}^2 t_{jl'}^2}/{U^2}$. Similar expression was also derived for one dimensional Heisenberg spin chain in terms of effective operators. \cite{Naef1998} The expression for $\alpha_{j,kl}$ implies that, in contrast to the case of the electric current density, 
loops are not needed to get finite contributions to $\tilde{{\bf I}}_{jk}^{(h)}$ (see Fig.~\ref{f1}). In other words, the effective heat current density operator is non-zero for a one-dimensional system with only nearest-neighbor hopping. This additional difference between the effective heat and electric current density operators arises from the fact that the net electric current 
$\langle \sum_{jk} \tilde{{\bf I}}_{jk}^{(c)} \rangle$ is always zero in the low-energy sector of  a Mott insulator (electrons are localized), while the net heat current
$\langle \sum_{jk} \tilde{{\bf I}}_{jk}^{(h)} \rangle$ can be finite (spin excitations can transport energy). 

At this point it is important to emphasize that  $\tilde{{\bf I}}_{jk}^{(c)}$ and $\tilde{{\bf I}}_{jk}^{(h)}$ cannot be obtained from the continuity equations for the effective operators:
$\partial_t {\tilde \rho}+\nabla\cdot {\tilde {\bf I}}^{(c)}$=0 and $\partial_t {\tilde \epsilon}+\nabla\cdot {\tilde {\bf I}}^{(h)}$=0. The reason is that the effective current density operators on adjacent bonds $jk$ and $kl$ have a common contribution if the hopping $t_{jl}$ is non-zero (triangular loop). The common contributions cancel out in the divergence of the effective current density operator. Therefore, knowing  $\nabla\cdot {\tilde {\bf I}}$ is not enough to obtain $ {\tilde {\bf I}}$.

\begin{figure}[t]
\includegraphics[angle=-90,width=\columnwidth]{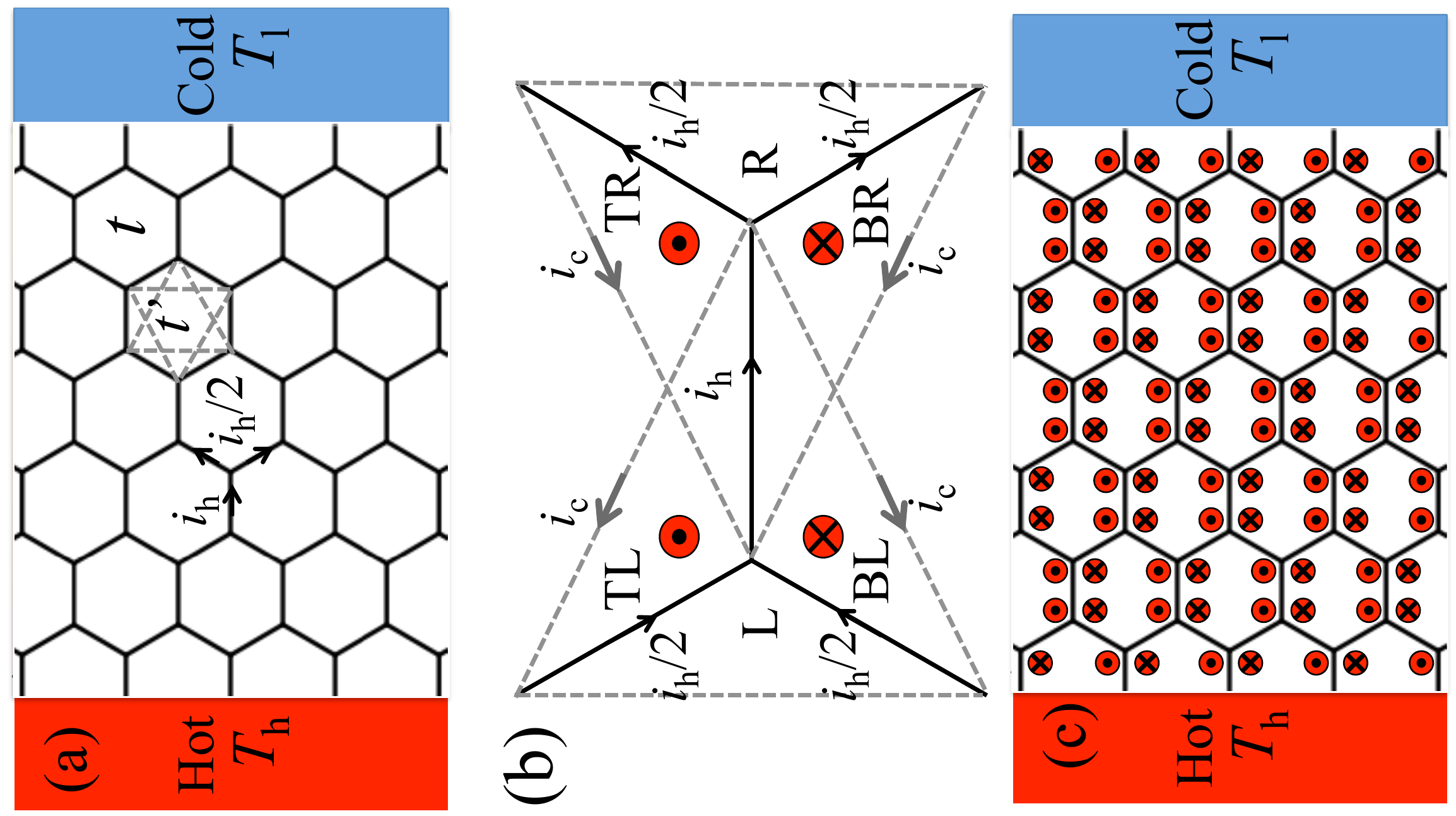}
 \caption{(color online) a) Hubbard model on a honeycomb lattice  with nearest  and next-nearest-neighbor hoppings, $t$ (full lines) and $t'$ (dashed lines),  and different temperatures $T_h$ and $T_l$ on both sides of the system. b) The heat current $i_h$ induces a finite scalar spin chirality with opposite signs on the top (TR and TL) and
bottom (BU and BL) triangles. Because $t'$ is finite, the local scalar spin chirality produces orbital currents [see Eq.\eqref{effcc}], which generate antiferromagnetically ordered orbital magnetic moments. c) Orientation of the orbital magnetic moments induced by the heat current. Circled dots (crosses) denote up (down) moments. }
  \label{f2}
\end{figure}

Our next goal is to demonstrate that the common nature of $\tilde{{\bf I}}_{jk}^{(c)}$ and $\tilde{{\bf I}}_{jk}^{(h)}$ (both are proportional to the local scalar spin chirality) leads to novel effect in Mott insulators on particular lattice. To illustrate this point we will assume that ${\cal H}$ is defined on a honeycomb lattice with only nearest and next nearest hopping amplitudes, $t$ and $t'$ respectively, with $t'\ll t$. We will also assume that both sides of the system are connected to different thermal baths with temperatures $T_h$ and $T_l$ [see Fig.~\ref{f2} (a)]. If the highest temperature, $T_h$, is much lower than the charge gap of the Mott insulator ($k_B T_h \ll U$) , we can use our low-energy effective model ${\tilde {\cal H}}$ and operators ${\tilde O}$ to describe the electronic properties of the system under consideration.

 The finite temperature difference, $\Delta T = T_h - T_l$, induces a  heat current density $\langle \tilde{{\bf I}}_{jk}^{(h)} \rangle = i_h {\hat x}$ on the horizontal bonds $jk$ and $\langle \tilde{{\bf I}}_{kl}^{(h)} \rangle = \pm i_h  {\hat e}_{\pm}/2$, on the oblique bonds $kl$, where ${\hat e}_{\pm} = {\hat x}/2 \pm \sqrt{3} {\hat y}/2$ with $\hat{x}$ ($\hat{y}$) the unit vector along the $x$ ($y$) direction. According to Eq.~\eqref{effhc}, this heat current density distribution must arise from a non-zero distribution of ${\chi}_{jkl}\equiv\langle \hat{\chi}_{jkl} \rangle$. We will adopt the convention that the three sites $jkl$ are oriented clockwise. Because the system is translationally invariant, there are only six types of triangles that are depicted in Fig.~\ref{f2} (b). The labels of the 
 six triangles, right (R), left (L), tip-right (TR), top-left (TL), bottom-right (BR) and bottom-left (BL),  are relative to a horizontal bond. The system remains translationally invariant  and preserves its mirror symmetry plane perpendicular to the ${\hat y}$-axis in presence of the uniform heat current density along ${\hat x}$. The other symmetry that survives is the product of a reflection in the plane perpendicular to the ${\hat x}$-axis  and time reversal. These remaining symmetries imply that the mean value of scalar spin chirality is zero for the R and L triangles, $\chi_{\mathrm{R,L}}=0$, while it is finite and of opposite signs for the the T and B triangles: $\chi_{\mathrm{TM}}=-\chi_{\mathrm{BM}}=\chi$, with $M=\mathrm{L,\ R}$. Knowing $\langle \hat{\chi}_{jkl} \rangle$ on each triangle, we can obtain the mean value of the heat current density on the horizontal bonds from Eq.~\eqref{effhc}, and the electric current density, $i_c$, on the 
 oblique dashed bonds shown in Fig.~\ref{f2} (b), from Eq.~\eqref{effcc}:
 \begin{equation}
 i_h = \frac{-32 t^4}{\hbar U^2} \chi, \;\;\; i_c = \frac{-24 e t^2 t'}{\hbar U^2} \chi.
 \end{equation}
Here $i_h$ does not depend on $t'$ because we are neglecting  contributions of order $t'^2/t^2$. 
The electric orbital current circulates around the top and bottom triangles and the corresponding orbital magnetic moments are
$
\mu = \pm   i_c A = \pm 3et' A  i_h / 4t^2,
$
 where the $+$ ($-$) sign holds for the top (bottom) triangles [see Fig.~\ref{f2} (c)] and $A$ is the area of a triangle. These orbital magnetic moments  are originated by the local scalar spin chirality induced by the heat current. 
 \begin{figure}[t]
\includegraphics[width=\columnwidth]{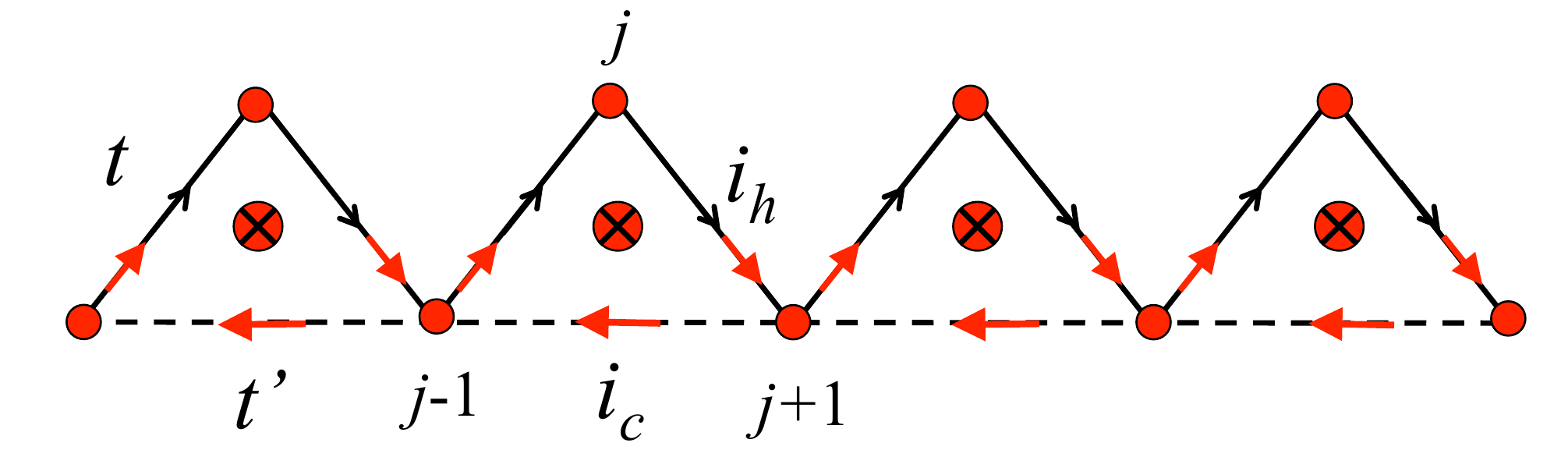}
 \caption{(color online) Sawtooth chain with hoppings $t$ and $t'$, where $t'\ll t$. Black arrows indicate the circulation of the hear current $i_h$, while red arrows denote the electric current. The  magnetic moments induced by orbital electric currents are indicated with crossed circles.}
  \label{f3}
\end{figure}
 
 An even simple example is provided by the Hubbard model defined on the sawtooth chain depicted in Fig.~\ref{f3}. The dominant contribution to the heat current comes from the nearest-neighbor hopping $t$ and it is 
$
 i_h =  -16 t^4 \chi_0 / \hbar U^2 .
$
Here $\chi_0\equiv\langle \mathbf{S}_{j+1}\cdot(\mathbf{S}_{j-1} \times \mathbf{S}_{j}) \rangle$ does not depend on the site index $j$ because the system is translationally invariant by one lattice parameter ($j \to j+1$ ) when $t'=0$ (we are neglecting contributions of order $t'^2/t^2\ll 1$). The finite $\chi_0$ leads to orbital currents 
\begin{equation}
 i_c =  \frac{3et'}{2t^2} i_h,
 \label{icih}
 \end{equation}
 that circulate around the triangles. In contrast to the previous case, the orbital currents are all oriented in the same direction (see Fig.~\ref{f3}), i.e., the heat current induces {\it orbital ferromagnetism} and the magnitude of each orbital moment is $\mu =  3et' i_h A/2t^2$. To understand the origin of this {\it uniform orbital magnetization} induced by a thermal current, it is convenient to go back to Eqs.~\ref{eq5} and \ref{heatc2} to notice that
\begin{equation}
 {\bf I}^{(h,U)}_{jk} = \frac{U}{2e} \delta \rho_{jk} {\bf I}^{(c)}_{jk},
 \label{new}
\end{equation}
 where  ${\bf I}^{(h,U)}_{jk}$ is the Coulomb contribution to the heat current density. Multiplying both sides of Eq.~\eqref{new} by $\delta \rho_{jk} {\bf R}_{jk}$, where ${\bf R}_{jk}=({\bf r}_j+{\bf r}_k)/2$ is the coordinate of the bond $jk$, we obtain
\begin{equation}
{\bf P}_{jk}\times {\bf I}^{(h,U)}_{jk} = \frac{U}{2e} \delta \rho^2_{jk} {\bf M}_{jk}.
\label{mte}
\end{equation}
Here ${\bf P}_{jk} = \delta \rho_{jk} {\bf R}_{jk}$ and ${\bf M}_{jk} =  {\bf R}_{jk} \times {\bf I}^{(c)}_{jk}$ are the electric and magnetic polarization densities. Equation~\eqref{mte} implies that a uniform thermal current density induces a net magnetization only if the Mott insulator has a net electric polarization:
$\langle {\bf M} \rangle \propto \langle {\bf P} \rangle \times \langle {\bf I}^{h} \rangle$, where ${\bf P} = \sum_{\langle jk\rangle} {\bf P}_{jk}$ and ${\bf M} = \sum_{\langle jk\rangle} {\bf M}_{jk}$ are the macroscopic electric and magnetic polarizations.  In other words, the Mott insulator must be ferroelectric or the lattice must break inversion symmetry, like  the sawtooth chain of Fig.~\ref{f3}, for the thermal current to induce a net orbital magnetization. Indeed, by taking mean values in  Eq.~\eqref{mte} and using that $| \langle {\bf P}_{jk} \rangle | \propto t^2 t'/U^3$ and $\langle \delta \rho^2_{jk}\rangle \propto t^4/U^4$ for $U \gg |t|$, we obtain 
$|\langle {\bf M}_{jk} \rangle | \propto e t' |\langle {\bf I}^{(h)} \rangle|/t^2$, in agreement with Eq.~\eqref{icih}.

For a magnetic contribution to the thermal conductivity of the order of 100 $\mathrm{W /(m\cdot K)}$ and an exchange constant of 1000 K, \cite{Hess07} the orbital moments induced by a thermal gradient of 10 $\mathrm{K/\mu m}$ are of order $10^{-4}\ \mu_B$, where $\mu_B$ is the Bohr magneton (we are assuming that $U/t\approx 10$). This magnetic moment corresponds to a magnetic field value at the center of each triangle of the order of 1G \footnote{ We note that this order of magnitude estimate is not modified by the contribution from currents in triangles which are different from the one that surrounds the point under consideration}. 
Although these are small magnetic moments, the magnitude of the effect should be significantly larger in the intermediate coupling regime because $| \langle {\bf P}_{jk} \rangle | / U \langle \delta \rho^2_{jk}\rangle$ remains of the same order but the electronic contribution to the thermal current becomes much larger.  Indeed, magnetoelectric effects measured in the 
intermediate-coupling organic Mott insulator $\kappa$-(BEDT-TTF)$_2$Cu$_2$(CN)$_3$ indicate a rather strong spin-charge coupling that is relevant for understanding
the low temperature properties of this spin liquid candidate~\cite{Poirier12}. 


In summary, we have derived the effective heat current operator for the strong-coupling limit of the half-filled Hubbard model and demonstrated that, like in the case of the electric current density, the leading order contribution is proportional to the scalar spin chirality. This common property of both current density operators is
dictated by symmetry considerations. The physical consequence of this commonality is a novel thermomagnetic effect: heat currents  induce orbital magnetic moments in frustrated Mott insulators. These  moments can  be measured with NMR if a large enough temperature gradient can be applied to the Mott insulator ( temperature gradients of $50 \ \mathrm{K/\mu m}$ can be applied to nano-devices~\cite{Slachter10}). Moreover, we have shown that the orbital moments produce a net magnetization, which is much easier to measure with conventional methods, if a net electric polarization is present. This spin-charge effect should be much stronger in the intermediate-coupling regime that is relevant for several frustrated Mott insulating materials. We emphasize that this thermomagnetic effect does not rely on any (low-energy) quasi-particle description, such as spin-waves for magnetically ordered states, and remains valid in the diffusive high temperature regime, i.e., for temperatures higher than the magnetic ordering temperature.

This work was carried out under the auspices of the NNSA of the US DoE at LANL under Contract No. DE-AC52-06NA25396, and was supported by the US Department of Energy, Office of Basic Energy Sciences, Division of Materials Sciences and Engineering.

%

\end{document}